\documentclass[showpacs,preprintnumbers,amsmath,amssymb,prb,twocolumn]{revtex4}

\usepackage{graphicx}
\usepackage{dcolumn}
\usepackage{bm}
\def\comment#1{}

\def\sigmab{{\mbox{\boldmath $\sigma$}}}

\def\slashchar#1{\setbox0=\hbox{$#1$}           
   \dimen0=\wd0                                 
   \setbox1=\hbox{/} \dimen1=\wd1               
   \ifdim\dimen0>\dimen1                        
      \rlap{\hbox to \dimen0{\hfil/\hfil}}      
      #1                                        
   \else                                        
      \rlap{\hbox to \dimen1{\hfil$#1$\hfil}}   
      /                                         
   \fi}                                         %

\begin{document}

\title{Compact quantum electrodynamics
in $2+1$ dimensions and spinon deconfinement: a renormalization group analysis}

\author{Flavio S. Nogueira}
\email{nogueira@physik.fu-berlin.de}
\affiliation{Institut f{\"u}r Theoretische Physik,
Freie Universit{\"a}t Berlin, Arnimallee 14, D-14195 Berlin, Germany}
\author{Hagen Kleinert}
\email{kleinert@physik.fu-berlin.de}
\affiliation{Institut f{\"u}r Theoretische Physik,
Freie Universit{\"a}t Berlin, Arnimallee 14, D-14195 Berlin, Germany}

\date{Received \today}

\begin{abstract}
We discuss compact $(2+1)$-dimensional Maxwell electrodynamics
 coupled to fermionic matter with $N$ replica.
For large enough $N$, the latter
corresponds to an effective theory for the nearest neighbor SU(N) Heisenberg
antiferromagnet, in which the fermions represent solitonic excitations
known as spinons. Here we show that the spinons are deconfined for $N>N_c=36$, thus leading to
an insulating state known as spin liquid. A previous analysis considerably underestimated the
value of $N_c$. We show further
that for $20<N\leq 36$ there can be either a
confined or a deconfined phase,
depending on the instanton density. For $N\leq 20$ only the confined phase exist.
For the physically relevant value $N=2$ we argue that no paramagnetic phase can emerge, since
chiral symmetry breaking would disrupt it. In such a case a spin liquid or
any other nontrivial paramagnetic state (for instance, a valence-bond solid) is only
possible if doping
or frustrating interactions are included.
\end{abstract}

\maketitle

\section{Introduction}

Quantum electrodynamics in D+1 spacetime dimensions (QED${}_{\rm D,1}$)
with $\rm D=1,2$ are useful
field-theoretic models in high-energy physics.
Phenomena like chiral symmetry breaking
and confinement are easier to understand in QED${}_{1,1}$ and QED${}_{2,1}$ than in QCD.
The simplest, exactly solvable model of this type
is spinor
quantum electrodynamics
in $1+1$ spacetime dimensions (QED${}_{1,1}$), the so-called Schwinger model,
 which exhibits
both chiral symmetry breaking and confinement.\cite{Schwinger}
Another example, relevant to the present paper, is the
$(2+1)$-dimensional
spinor quantum electrodynamics,
QED${}_{2,1}$, in the form introduced
by Pisarski some time ago.\cite{Pisarski} This model is known to exhibit spontaneous
chiral symmetry breaking.\cite{Appelq-1,Appelq-2} An interesting aspect of
this model is its applicability to
condensed matter physics where it appears in
different contexts, especially in the study of
high-$T_c$ superconductors and Mott insulators.\cite{Affleck,Kim,Rantner,Tesanovic,Herbut2}
In the study of Mott insulators, it arises as an effective theory of the
so-called {\em spin liquids\/}, which are Mott insulators without any broken symmetry. In this case the Dirac fermions
represent the so-called {\em spinons\/},
soliton-like excitations carrying spin degrees of freedom but no charge.
A good name for this type of QED is
{\it quantum spinodynamics} (QSD), since it is actually a quantum field theory of spinons.
The theory
can be derived for
Mott insulators, and it is found that
the abelian gauge field coupling to
the
spinons is compact. This follows immediately by accounting for the fluctuations
around mean-field theories of resonating valence bonds (RVB)\cite{Anderson} states which
have a local U(1) gauge freedom
in which the phase angle is defined only modulo $2\pi$.
These mean-field theories are derived from the strong-coupling limit
of
the Hubbard model, the so-called Heisenberg-Hubbard model:
\begin{equation}
H=J\sum_{\langle i,j\rangle}{\bf S}_i\cdot{\bf S}_j,
\label{@HEI}\end{equation}
where
${\bf S}_i$ are spin-$1/2$ operators
formed from Fermi fields  $f_{i\beta}$
subjected to the
local constraint $f_{i\sigma}^\dagger f_{i\sigma}=1$ as
${\bf S}_i=(1/2)f_{i\alpha}^\dagger\sigmab_{\alpha\beta}f_{i\beta}$,
where $\sigmab$
are the Pauli matrices
$\sigmab\equiv(\sigma_1,\sigma_2,\sigma_3)$
(throughout
this paper summation over repeated greek indices is implied).
The sum in (\ref{@HEI}) runs over nearest neighbor
pairs of sites in a square lattice,  and
$J$ is related to
the original  parameters of hopping and interaction energies
$t$ and $U$ of the Hubbard model by
$J=4t^2/U$.
Different
composite-field theories
 are known
to represent the same quantum states of the model.
This follows from the fact that the {\it local} gauge
symmetry of the Heisenberg-Hubbard model is actually SU(2).\cite{Hsu}
Thus, the composite link fields
\begin{equation}
\Delta_{ij}^*=\langle f_{i\uparrow}^\dagger f_{j\downarrow}^\dagger-
f_{i\downarrow}^\dagger f_{j\uparrow}^\dagger\rangle,
\end{equation}
and
\begin{equation}
\label{AM-OP}
\chi_{ij}=\langle f_{i\sigma}^\dagger f_{j\sigma}\rangle,
\end{equation}
where $(i,j)$ are nearest neighbors, which are
obtained from different Hubbard-Stratonovich decouplings of the Heisenberg-Hubbard model, describe
the {\it same} physics when associated with the most stable ground state of the
corresponding mean-field theory. They are connected through a SU(2) gauge transformation.\cite{Hsu}

The phase fluctuations of either
composite field
features a lattice gauge field $A_{ij}$.
These are obviously
of the compact  U(1) type.
Both fields
together transform into each other
by  local SU(2) transformations, and theories utilizing this symmetry
have
been studied in the past,\cite{Wen,Lee-Wen}
with some
advantages over
compact
U(1) theories, especially if one is interested in studying the
phase structure of
high-$T_c$ superconductors.\cite{Lee-Wen,Wen-RMP,Ki-Seok-Kim} A local SU(2) gauge theory also emerges in 
the study of frustrated Heisenberg antiferromagnets.\cite{Mudry-2d,Mudry-1d} 
The existence of the larger SU(2) symmetry
ensures the compactness
of the Abelian theories even in the continuum limit,
 since
the U(1) group is a subgroup of SU(2).\cite{Polyakov}

A controlled study of U(1) spin liquids was initiated some time ago by
Affleck and Marston\cite{Affleck} using the composite field (\ref{AM-OP}).
To have a small expansion parameter,
 they generalized the {\it global} symmetry SU(2) to SU(N)
and solved the model in the large
$N$ limit. This was done in the so-called self-conjugate representation of
SU(N),\cite{Arovas,ParaMarston} where the spin operators are given by
$S_{i,\alpha\beta}= f_{i\alpha}^\dagger f_{i\beta}-\delta_{\alpha\beta}/2$
and fulfill the local constraint
$f_{i\alpha}^\dagger f_{i\alpha}=N/2$. These operators have zero trace:
${\rm Tr}(S_{i,\alpha\beta})=0$.
If the partition function is represented as a functional integral
over the action
associated with the Hamiltonian
(\ref{@HEI})
we may calculate a mean-field approximation
from the
saddle point approximation.
The result
which  preserves all the
lattice symmetries is the so-called
$\pi$-flux phase,\cite{Affleck} whose spectrum of elementary excitations is
given by
\begin{equation}
E_{\bf k}=2|\chi_0|\sqrt{\cos^2k_x+\cos^2k_y},
\end{equation}
where $|\chi_0|$ is the mean-field amplitude of $\chi_{ij}$. The excitations
around the Fermi points $\pm(\pi/2,\pi/2)$ can be at low energies
represented as four-component Dirac fermions.
The phases
of
$\chi_{ij}$ fluctuate strongly
and form a link gauge field
 $A_{ij}$.
It is not difficult to show that
the effective low-energy Lagrangian in imaginary time
has the form\cite{Note1}
\begin{equation}
\label{L-F}
{\cal L}=\frac{1}{4e_0^2}F_{\mu\nu}^2+
\sum_{a=1}^N\bar \psi_a(\slashchar{\partial}+i\slashchar{A})\psi_a,
\end{equation}
where $F_{\mu\nu}=\partial_\mu A_\nu-\partial_\nu A_\mu$ and
we have used the
usual Feynman slash notation $\slashchar{a}\equiv\gamma_\mu a_\mu$,
with $\gamma_\mu$
being $4\times 4$ $\gamma$ matrices. In the above continuum notation the
compactness
of the gauge field is not apparent. As it stands, the above Lagrangian is
just the
above-mentioned massless QED$_{2,1}$, which is a well studied model. We
shall reserve
the abreviation QSD for the compact version of 
${\rm QED}$$_{2,1}$, whose field theory will be
discussed at lenght in Sect. III.

In QSD the spinons play a role similar to quarks in QCD. Indeed,
compactness of
the gauge field leads to spinon confinement if $N$ is not large enough.\cite{Kim,Hermele,KNS,IL}
Note, however, an important difference. In QCD
the gluon is introduced ad hoc to generate the coupling between the quarks.
The ``gluon'' in QSD, on the other hand, has a clear origin: it is spontaneously
generated by the phase fluctuations of the link field (\ref{AM-OP}), which in turn is a composite field
made of lattice fermions. This unique feature of QSD led Wen\cite{Wen-Photon}
to propose a similar mechanism in 3+1 dimensions to explain the origin of gauge bosons.

Whether spinons in QSD deconfine or not was
for some time matter of controversy.
\cite{KNS,Nayak,KNS1,Herbut1,Herbut3,Kragset} The controversy seems now to
be solved,\cite{Hermele,NK} at least at a qualitative level. The result is
that spinons deconfine for large enough $N$,\cite{Hermele} and this guarantees
the stability of the large-$N$ spin liquid. The aim of this paper is to
further improve our understanding of spinon deconfinement, in particular by
a more quantitative analysis based on the renormalization group (RG). We start in Sect. II
with a discussion of the model in the absence of matter, which
just corresponds to the
compact
version of Maxwell electrodynamics in $2+1$ dimensions
($\overline{\rm QED}$$_{2,1}$).\cite{Polyakov} The
equivalence of this model with a three-dimensional Coulomb gas of instantons allows us
to make a relatively simple field theoretical analysis, based on the equivalence of a
a Coulomb gas with the sine-Gordon theory.
In Sect. II we compute the one-loop effective potential of the
model in terms of a scalar field $\varphi(x)$ whose correlation function
$\langle\varphi(x)\varphi(x')\rangle$ gives a direct measure of the interaction between
instantons in three-dimensional spacetime. The calculation of the effective potential
leads to an estimate of the instanton mass (which is different from the
mass of the scalar field). We also derive the RG equations for the model. The
reulting effective potential and the RG equations confirm in a
physically appealing way the well-known result\cite{Polyakov,Goepfert} that in absence of matter the
(2+1)-dimensional compact gauge theory permanently confine test charges. This result contrasts
with the one in 3+1 dimensions, where a deconfined phase has been proven to
exist.\cite{Peskin,Guth,Klei-ColWein}

Section III is where we start with the analysis of compact QED$_{2,1}$ or QSD. Here is
very difficult to write a field theory in terms of the original variables. It will be necessary,
just like in Sect. II, to work with a field theory describing the dynamics
of the instantons. We still have a gas of instantons, but it is no
longer of the Coulomb type, since the
interaction between the instantons is modified by the vacuum polarization. The RG equations
are obtained in essentially the same way as in Sect. II, except that the effects of the
vacuum polarization are taken into account. These change the phase structure of the theory in
an essential way. By considering a theory with $N$ replica of Dirac fermions, we show that
spinons deconfine for $N>N_c=36$. This is a considerable numerical change with respect with our
previous calculation of $N_c$,\cite{NK} where a value smaller by a factor $\pi^3$ was
found due to a wrong counting of $\pi$ factors in the RG calculation. Apart from this
mistake, the analysis in Ref. \onlinecite{NK} is correct, in the
sense that a critical value of $N$ is predicted above which deconfinement occurs.
Here we elaborate further the nature of the confined phase.
In particular we show that confinement is the only possibility for
$N\leq 20$ while deconfinement certainly occurs only for $N>36$. In the
region $20<N\leq 36$ either phase is possible, depending on the instanton density.
The universal properties of QSD are encoded in the coefficient of the $-1/R$ contribution
to the interspinon potential. In the deconfined phase this coefficient is given by
the fixed point in the renormalization flow
derived in the $1/N$ expansion. However, in the confined phase the
$1/N$ expansion does not hold and an expansion in the fugacity is made instead. Spinon
deconfinement implies the stability of the spin liquid. The experimentally relevant
case is the $N=2$ one, which corresponds to the regime where the spin liquid is
unstable, since the spinons are confined in this case. The nature of the confined phase
for $N=2$ is discussed in Sect. IV, where the important role played by
chiral symmetry breaking is
emphasized.

\section{Compact Maxwell electrodynamics in 2+1 dimensions}

The $(2+1)$-dimensional compact electrodynamics studied by Polyakov,\cite{Polyakov}
abreviated here as $\overline{\rm QED}_{2,1}$, is
actually Maxwell theory in $2+1$ dimensions in which the gauge group U(1) is
made compact. The model was originally motivated by the spontaneous
symmetry breaking pattern of the Georgi-Glashow model in $2+1$ dimensions.
After spontaneous symmetry breaking of the SU(2) group one is left with a residual
U(1) group which is compact, since it is a subgroup of SU(2).
Another place where this theory arises naturally is in lattice gauge theory.\cite{Polyakov-book}

It is well known that $\overline{\rm QED}_{2,1}$ is
equivalent via duality to a three-dimensional
Coulomb gas of instantons,\cite{Polyakov}
whose energy interaction is given by
\begin{equation}
\label{moninterac}
E_I=\frac{2\pi^2}{e_0^2}\sum_{i,j}\frac{q_iq_j}{4\pi|x_i-x_j|},
\end{equation}
where $e_0$ is the bare gauge coupling and $q_i=\pm q$ with $q\in\mathbb{N}$
are the instanton charges.

In two dimensions the Coulomb gas corresponds to a theory dual to the
two-dimensional XY model \cite{Jose,Kleinert-GFCM-1} or, equivalently, a two-dimensional
classical superfluid.\cite{KT} The Coulomb interaction between
the charges is in this case proportional to $\ln|x_i-x_j|$. In the
context of two-dimensional superfluids, the charges in the Coulomb
gas are interpreted as vortices in two dimensions. The two-dimensional
Coulomb gas is known to undergo a vortex-antivortex pair
unbiding phase transition, the cellebrated Kosterlitz-Thouless (KT)
phase transition.\cite{KT} In the language of the Coulomb gas we have a
low-temperature dielectric phase separated from the
high-temperature ``metallic'' plasma phase by a phase transition without
breaking the U(1) symmetry of the original superfluid or XY system,
in agreement with the Mermin-Wagner theorem.\cite{MW} In three dimensions,
on the other hand, no phase transition occurs in the Coulomb gas and
the system remains in the plasma phase. In this case the Debye-H\"uckel (DH) mean-field
theory is essentially correct and the interaction is screened in
such a way that the excitations are always gapped. This can be
conveniently expressed in field theory language by means of the
sine-Gordon representation of the Coulomb gas. In the context of
$\overline{\rm QED}_{2,1}$ the corresponding sine-Gordon theory reads
\begin{equation}
\label{SG}
{\cal L}_{\rm SG}=\frac{1}{2}
(\partial_\mu\varphi)^2-2z_0\cos\left(\frac{2\pi}{e_0}\varphi\right).
\end{equation}
The above Lagrangian corresponds to the field theory model dual to
$\overline{\rm QED}_{2,1}$. The parameter $z_0$ is the bare fugacity of the Coulomb
gas. The DH theory amounts to a
Gaussian approximation to the above Lagrangian. This leads to a
mass $M_0=2\pi\sqrt{2z_0}/e_0$ for the scalar field $\varphi$. This
behavior of the dual model implies a corresponding mass gap in the
magnetic field correlation function.\cite{Polyakov}
Fluctuation corrections to the DH approximation essentially
do not change this result. To see this, let us
compute the one-loop effective potential.
This can be easily obtained by standard methods.\cite{Jackiw}
At one-loop order it is more easily obtained by writing $\varphi=\bar \varphi+\delta\varphi$,
where $\bar \varphi$ is a constant background field while $\delta\varphi$ represents a small
fluctuation around it. By integrating out the Gaussian fluctuations, the one-loop effective
potential is obtained:
\begin{equation}
V_{\rm eff}(\bar \varphi)=-2z_0\cos\left(\frac{2\pi}{e_0}\bar \varphi\right)
-\frac{2\pi^2}{3}\left[\frac{z_0}{e_0^2}\cos\left(\frac{2\pi}{e_0}\bar \varphi\right)\right]^{3/2},
\end{equation}
where a counterterm proportional to $\cos(2\pi\varphi/e_0)$ was used to trivially
subtract the contribution $(2z_0\Lambda/e_0^2)\cos(2\pi\varphi/e_0)$, with
$\Lambda$ being an ultraviolet cutoff.
As with Eq. (\ref{SG}), the obtained effective potential implies a degenerate vacuum at
$\bar \varphi_n=ne_0$, $n\in\mathbb{Z}$, whose energy density is given by
$E_0=-2z_0-(2\pi^2/3)(z_0/e_0^2)^{3/2}$, which is always negative, just
as the energy of the vacuum without the quantum corrections.
From this we conclude that the instantons of
this theory are always massive.

In order to better understand the meaning of this statement,
let us compare the above effective potential with the one of a (1+1)-dimensional sine-Gordon theory.
In this case the corresponding Euclidean theory is the dual field theory of a
two-dimensional superfluid,\cite{Jose} which is known to undergo a KT phase transition.\cite{KT}
The one-loop effective potential reads:\cite{Joseph}
\begin{eqnarray}
&V_{\rm eff}(\bar \varphi)=-2z\cos\left(2\pi\sqrt{K}\bar \varphi\right)
\nonumber\\
&-\pi Kz\cos\left(2\pi\sqrt{K}\bar \varphi\right)
\ln\left[\frac{4\pi^2Kz}{\Lambda^2}\cos\left(2\pi\sqrt{K}\bar \varphi\right)\right],
\end{eqnarray}
where $K$ is the superfluid stiffness and $\Lambda$ an ultraviolet cutoff. In order
to remove the cutoff we add a counterterm proportional to $\cos\left(2\pi\sqrt{K}\bar \varphi\right)$
and impose the renormalization condition $V_{\rm eff}''(0)=8\pi^2Kz\equiv M_\varphi^2$, which just fixes the
renormalized mass of the scalar field in such a way as to have the 
same form as the one obtained from the
DH theory. The result is
\begin{eqnarray}
&V_{\rm eff}(\bar \varphi)=-z(2-\pi K)\cos\left(2\pi\sqrt{K}\bar \varphi\right)
\nonumber\\
&-\pi Kz\cos\left(2\pi\sqrt{K}\bar \varphi\right)
\ln\left[\cos\left(2\pi\sqrt{K}\bar \varphi\right)\right].
\end{eqnarray}
The ground state energy density is now $E_0=z(\pi K-2)$. This changes sign at
$K_c=2/\pi$, precisely at the critical KT stiffness value.\cite{Jose} In this
case we can use the results of Ref. \onlinecite{Neveu} to obtain the
soliton mass as
\begin{equation}
M_{\rm sol}=\frac{2}{\pi}\sqrt{\frac{2z}{K}}(2-\pi K).
\end{equation}
For $K\geq K_c$, which
corresponds to low temperatures in the KT theory, the soliton mass vanishes while
the energy becomes positive. For $K<K_c$, on the other hand, we can write
\begin{equation}
\label{Msol-E0}
M_{\rm sol}=-\frac{2}{\pi}\sqrt{\frac{2}{zK}}E_0=-\frac{8E_0}{M_\varphi}.
\end{equation}
Note that
while the soliton mass vanishes at $K_c$, the scalar field mass does not. This is
very important, since in the (2+1)-dimensional case the mass of the scalar field
also does not vanish for any $e_0^2$,
but there the same is true of the mass of instanton excitations. By analogy to
Eq. (\ref{Msol-E0}), we can infer that the instanton mass in CMT$_3$ should be
given by
\begin{equation}
M_{\rm inst}\propto -\frac{E_0}{M_0}
=\frac{e_0}{\pi\sqrt{2z_0}}\left[z_0+\frac{\pi^2}{3}\left(\frac{z_0}{e_0^2}\right)^{3/2}\right],
\end{equation}
which never vanishes in contrast to Eq. (\ref{Msol-E0}).

The above results are made more transparent when we consider the RG equations of
$\overline{\rm QED}_3$, which are equivalent to the RG equations of the three-dimensional
Coulomb gas. To lowest order this may be achieved by means of a
scale-dependent DH approximation, in the same spirit of the analysis of the
two-dimensional Coulomb gas made by Young.\cite{Young} The RG equations for the
three-dimensional Coulomb gas were originally derived by Kosterlitz,\cite{Kosterlitz}
using the so-called poor-man scaling approach. 
As shall we see, for our purpose it is better to use Young's
approach,\cite{Young} which we generalize to the higher-dimensional case. The
derivation is given in Appendix A for the case of a $d$-dimensional Coulomb gas.
Again, it will be useful to compare the RG equations for the KT case with those for
$\overline{\rm QED}_{2,1}$. Setting $d=2$ in Eqs. (\ref{RG-K-new}) and (\ref{dz-new}) of Appendix A, we
obtain the well known RG equations for a two-dimensional superfluid where $\kappa=K$:
\begin{equation}
\label{KT-K}
\frac{dK^{-1}}{dl}=y^2,
\end{equation}
\begin{equation}
\label{KT-y}
\frac{dy}{dl}=(2-\pi K)y.
\end{equation}
From these equations we can see better why our approach of $K_c$ from
above drives the phase transition. In this case the sign on the right-hand-side of Eq.
(\ref{KT-y}) is negative, such that the system will flow to a regime of zero fugacity, leading to
a line of fixed points. At this line the mass of the scalar field
vanishes. If we define a dimensionless soliton mass through
$m_{\rm sol}\equiv M_{\rm sol}/\Lambda$, we obtain
\begin{equation}
\label{flow-msol}
\frac{dm_{\rm sol}^2}{dl}=\frac{2(2/\pi)^2}{K}
[(2-\pi K)^2+2y^2K].
\end{equation}
The fixed point of Eq. (\ref{flow-msol}) corresponds precisely to the KT
critical point. In this equation it should be understood that for $y=0$ and $K\geq K_c$
its right-hand side vanishes, since $M_{\rm sol}$ is zero for all $z\geq 0$ and
$K\geq K_c$.

Now let us see what happens in $\overline{\rm QED}_3$.
In this case  we set $K_0=a=1/e_0^2$ in the equations
of Appendix A, where $a$ is a short-distance cutoff, and
define the dimensionless gauge coupling
$f\equiv K_0/K=1/\kappa=e^2/e_0^2$, such that
\begin{equation}
f(l)=e_0^2ae^l\varepsilon(ae^l)=e^l\varepsilon(ae^l),
\end{equation}
with $l\equiv\ln(r/a)$ being a logarithmic length scale. Thus,
$f(l)e^{-l}$ is just given by the screening constant (``dielectric'' constant) of the
Coulomb gas of instantons.
In this way we obtain from Eqs. (\ref{RG-K-new}) and (\ref{dz-new})
the RG equations for the $\overline{\rm QED}_3$:
\begin{eqnarray}
\label{betaf-Polyakov}
\frac{df}{dl}&=&y^2+f,\\
\label{betay-Polyakov}
\frac{dy}{dl}&=&\left(3-\frac{\pi}{2f}\right)y,
\end{eqnarray}
Note that
the coefficient of the term $1/f$ in
Eq. (\ref{betay-Polyakov}) is different from the one obtained by us in Ref. \onlinecite{NK},
where factors of $\pi$ were overcounted.
It is easy to see that, in contrast with
Eqs. (\ref{KT-K}) and (\ref{KT-y}), the RG flow pattern does not exhibit any fixed point, giving
a further confirmation that the three-dimensional sine-Gordon model does not undergo
any phase transition. In other words, the excitations are always gapped.

\section{Compact QED in 2+1 dimensions (Quantum spinodynamics)}


When fermionic matter is present, the energy of the instanton gas (\ref{moninterac})
is changed due to
the vacuum polarization. The effective bare gauge coupling is modified to
$e^2(p)=Z_A(p)e_0^2$, where $Z_A(p)$ is the wave function renormalization of
the gauge field in the noncompact theory. We have
\begin{equation}
\label{ZA}
Z_A(p)=\frac{1}{1+\Pi(p)},
\end{equation}
where $\Pi(p)$ is the vacuum polarization. Thus, the energy of the instanton gas becomes
\begin{equation}
E_I=-\frac{1}{2}\sum_{i,j}U_0(x_i-x_j)q_iq_j,
\end{equation}
where
\begin{equation}
\label{bare-U-1}
U_0(x)=-\frac{4\pi^2}{e_0^2}\int\frac{d^3p}{(2\pi)^3}
\frac{e^{ip\cdot x}}{p^2}[1+\Pi(p)],
\end{equation}
is the interaction between two instantons of opposite charge.

Let us emphasize the dual nature of the above potential by writing the
{\it static} semiclassical interspinon potential between two opposite test spinon charges:
\begin{equation}
V(R)=-e_0^2\int\frac{d^2p}{(2\pi)^2}\frac{e^{i{\bf p}\cdot {\bf x}}}{{\bf p}^2[1+\Pi({\bf p})]},
\end{equation}
where $R\equiv|{\bf x}|$.
We see that there is a manifest duality between ``electric'' and instanton
charges even after the vacuum polarization is included. Indeed, from Eq. (\ref{bare-U-1})
the effective squared instanton charge is given by $\tilde e^2(p)=4\pi^2/e^2(p)$, 
thus verifying the Dirac duality relation $e(p)\tilde e(p)=2\pi$ between the effective charges.

At one-loop order we have $\Pi(p)=Ne_0^2/8p$, implying that at large distances the
potential becomes
\begin{equation}
\label{pot-nc}
V(R)=-\frac{4}{\pi NR}+{\cal O}\left(\frac{1}{R^2}\right),
\end{equation}
instead of having the logarithmic behavior expected classically. Thus, quantum fluctuations
lead in two spatial dimensions to a large distance behavior of the potential similar
to the one of electrodynamics in three spatial dimensions. Interestingly, the above power is
$1/R$ for all dimensions $d\in(2,4)$ and the coefficient in front of the factor $1/R$ is
universal,\cite{NK} a situation similar to the one obtained from string models for the
interquark potential.\cite{Luescher}

Since the energy of the instanton gas is changed by the presence of matter,
the sine-Gordon theory (\ref{SG}) must be changed correspondingly.\cite{KNS,KNS1,Herbut1}
It is
easy to see that the field theory of the instanton gas
is now given by
\begin{equation}
\label{SG-new}
{\cal L}=\frac{e_0^2}{8\pi^2}(\partial_\mu\varphi)Z_A(\sqrt{-\partial^2})
(\partial_\mu\varphi)-2z_0\cos\varphi,
\end{equation}
where the symbol $Z_A(\sqrt{-\partial^2})$ has an operator meaning such that in momentum
space the gradient energy reads simply $(e_0^2/8\pi^2)Z_A(p)p^2\varphi(p)\varphi(-p)$.
Thus, the Lagrangian
in Eq. (\ref{SG-new}) corresponds to that in Eq. (\ref{SG}) with the bare
squared charge $e_0^2$ replaced by an effective one.
Unlike in $\overline{\rm QED}_{2,1}$ discussed in the previous Section, Eq.~(\ref{SG-new}) does not
provide an exact dual representation of the theory, since its derivation is
obtained in a harmonic approximation to the gauge field fluctuations.
However, to the order to which our RG equations were
calculated it provides an accurate dual field theory representation.

After evaluating the momentum integral in (\ref{bare-U-1}) using the
short-distance cutoff $a=1/e_0^2$ and an one-loop approximation to $\Pi(p)$, we
obtain the following interaction between two oppositely charged instantons:
\begin{eqnarray}
\label{bareU-new}
U_0(r)&=&-\frac{\pi}{e_0^2}\left[\frac{1}{r}-\frac{1}{a}-\frac{Ne_0^2}
{4\pi}\ln\left(\frac{r}{a}\right)\right]
\nonumber\\
&=&-\frac{\pi}{e_0^2r}+\frac{N}{4}\ln(e_0^2r)+\pi,
\end{eqnarray}
where $r\equiv|x|$. Let us compute the pair
susceptibility at large distances
using the above potential. This is obtained from the $r\to\infty$ limit
of Eq. (\ref{suscep}) of
Appendix A for $d=3$, and inserting the Boltzmann factor $n(r)\approx z_0^2 e^{-U_0(r)}$.
The pair susceptibility obtained in this way gives a measure of $\langle r^2\rangle$ and
is given by
\begin{eqnarray}
\chi_0^*&=&\lim_{r\to\infty}\chi_0(r)
=\frac{16\pi^3z_0^2}{3e_0^2}\int_a^\infty d\rho\rho^4e^{-U_0(\rho)},
\end{eqnarray}
where the subscript in
$\chi$ indicates that the calculation is done with the potential
$U_0(r)$. By introducing the dimensionless variable $u=e_0^2\rho/\pi$, we obtain
\begin{equation}
\chi_0^*=\frac{16\pi^{8-N/4}e^{-\pi}z_0^2}{3(e_0^2)^6}\int_{1/\pi}^\infty duu^{4-N/4}e^{1/u}.
\label{chi0}
\end{equation}
The integral
converges only for $N>20$. This leads to
$\varepsilon_0^*=1+4\pi\chi_0^*>1$ signaling a ``dielectric'' phase
for the instanton gas. Duality implies that the ``dielectric'' instanton phase corresponds
to a ``metallic'' phase for the spinons, i.e., the spinons are deconfined and we have
a spin liquid. On the other hand, for 
$N\leq 20$ the instantons are in the ``metallic'' phase, which corresponds
by duality to confined spinons. As we shall see, it is not quite accurate to say that spinons
deconfine for $N>20$. Actually it will be shown later that in the interval
$20<N\leq 36$ both confinement and deconfinement phases may occur. A stable deconfined phase
occurs only for $N>36$.

The arguments of the preceding paragraph are better understood by a RG analysis.
We shall derive
the RG equations using the method  of Appendix A, with the $U_0(r)$ given in Eq.
(\ref{bareU-new})
replacing the one in Eq. (\ref{bareU}).
The dielectric constant $\varepsilon(r)$ of the
instanton gas as a
function of the length scale $r$ has the same form as in Appendix A, except
that the renormalized
interaction between the instantons is in this case given by
\begin{equation}
\label{U}
U(r)=U(a)+\frac{\pi}{e_0^2}\int_a^r\frac{ds}{\varepsilon(s)s^2}\left
(1+\frac{Ne_0^2}{4\pi}s\right),
\end{equation}
which implies a renormalized dimensionless gauge coupling
\begin{equation}
\label{f}
f(l)=\frac{e^l\varepsilon(ae^l)}{1+\frac{N}{4\pi}e^l}.
\end{equation}
The first RG equation follows immediately from Eq. (\ref{f}):
\begin{equation}
\label{betaf}
\frac{df}{dl}=f+y^2-\frac{N}{4\pi}\frac{f^2}{\varepsilon},
\end{equation}
where we have defined
\begin{equation}
\label{y2}
y^2(l)=\frac{64\pi^4}{3}\frac{a^6z_0^2e^{6l-U(ae^l)}}{1+\frac{N}{4\pi}e^l}.
\end{equation}
The third term in Eq. (\ref{betaf}) represents the correction to Eq. (\ref{betaf-Polyakov})
due to the presence of fermionic matter. For $y=0$ the  instantons are
suppressed and $\varepsilon=1$, such that
Eq. (\ref{betaf}) reduces to the $\beta$ function of noncompact QED$_{2,1}$.
In such a situation we can
rewrite the potential (\ref{pot-nc}) in terms of the fixed point $f_*=4\pi/N$
of the noncompact theory:
\begin{equation}
\label{pot-nc-new}
V(R)=-\frac{f_*}{\pi^2 R}+{\cal O}\left(\frac{1}{R^2}\right).
\end{equation}

By differentiating Eq. (\ref{y2}) with respect to $l$, we obtain
\begin{equation}
\label{betay}
\frac{dy}{dl}=y\left(3-\frac{\pi}{2f}-\frac{N}{8\pi}\frac{f}{\varepsilon}\right),
\end{equation}
such that the correction to Eq. (\ref{betay-Polyakov}) due to fermionic matter is also obtained.
The flow of the screening constant of the instanton gas is given by
\begin{equation}
\label{betaeps}
\frac{d\varepsilon}{dl}=\frac{\varepsilon y^2}{f}.
\end{equation}
This follows from
\begin{equation}
\frac{d\varepsilon}{dr}=\frac{16\pi^3}{3}\frac{z_0^2a^4}{e_0^2}e^{4l-U(ae^l)},
\end{equation}
which is easily derived using Eqs. (\ref{diel-const}) and (\ref{suscep}) of Appendix A.

The last term in Eq. (\ref{betay}) is crucial for understanding
QSD from a RG point of view. This term was overlooked before\cite{Hermele,KNS,IL,Herbut1,Herbut3} and led to an
incomplete physical picture of the phase structure of QSD. In order to emphasize
the importance of this term, let us neglect it for a moment.
Considering $f_*=4\pi/N$ we find a
line of fixed points for zero fugacity as $N$ is varied.\cite{IL,KNS}
From this it was concluded\cite{IL,KNS} that QSD undergoes a KT-like deconfinement
transition in three spacetime dimensions. In this deconfinement
scenario the critical value of $N$ above which the spinons deconfine is $N_c=24$.
The reason why this term was overlooked before is that
it was generally assumed that the logarithm term in Eq. (\ref{bareU-new}) dominates
over the 3D Coulomb interaction at large distances. In other words, the large distance
limit was being taken at a too early stage. We may argue that if one
has a logarithmic gas of instantons in 3D it should not be particularly surprising that
a 3D KT-like transition emerges. However, to really prove such a statement is far
from being trivial and has led to recent controversies.\cite{Herbut1,Herbut3,Kragset}
Monte Carlo simulations\cite{Kragset} seem to support the scenario of a
KT-like transition for the 3D logarithmic gas. However,
it is now known that such a transition {\it does not}
describe correctly the deconfinement transition in QSD.\cite{Hermele,NK} Here
we shall give a more complete explanation of the results obtained
in Refs. \onlinecite{Hermele} and \onlinecite{NK}.

In Ref. \onlinecite{Hermele} it was shown that
the spinons deconfine for a large but finite $N$, but the critical
value of $N$ above which it happens was not calculated.
This value was estimated in our paper\cite{NK} to be
$N_c=36/\pi^3$. From the more thorough analysis
performed here we shall see that this estimate lies too low.
One trivial reason for this is an overcounting
of $\pi$ factors in the derivation of the RG equations, as was mentioned after
Eqs. (\ref{betaf-Polyakov}) and (\ref{betay-Polyakov}). But there are other aspects
of the RG flow of QSD which were not considered in Refs. \onlinecite{Hermele} and
\onlinecite{NK} and will now be included.

From Eqs. (\ref{betaf}), (\ref{betay}), and (\ref{betaeps}), we see that there is
apparently
for all $N\geq 20$ a {\it fixed line} at zero fugacity and
$f_*=4\pi\varepsilon_*/N$, $1\leq\varepsilon_*\leq N/20$.
Note that $\varepsilon\geq 1$, since $\chi\geq 0$.
However, it is clear that for vanishing fugacity we must
have $\varepsilon_*=1$, so that $N=20$ and
the fixed line actually collapses to
a fixed point with $f_*=\pi/5$. It is reasonable to assume that the RG functions are given
by expansions in the fugacity. For $y=0$ we have $\varepsilon(l)=1$ {\it for all} $l$,
such that Eq. (\ref{betaf}) reduces to the RG $\beta$ function of noncompact QED$_3$. Up to terms
of order $y^2$ we can legitimately approximate $f^2/\varepsilon\approx f^2$ and
$yf/\varepsilon\approx yf$ in Eqs. (\ref{betaf}) and (\ref{betay}), respectively. In this way
Eqs. (\ref{betaf}) and (\ref{betay}) decouple from Eq. (\ref{betaeps}) and have the
following fixed points at nonzero fugacity:
\begin{eqnarray}
\label{f+-}
f_\pm&=&\frac{2\pi}{N}\left(6\pm\sqrt{36-N}\right),\\
\label{y+-}
y_\pm^2&=&\frac{10\pi}{N}
\left(6-\frac{N}{10}\pm\sqrt{36-N}\right).
\end{eqnarray}
We confirm once more the fixed point with zero fugacity at $N=20$,
$y_-=0$ and $f_-=\pi/5$. The above fixed points do not exist if $N>36$, in which case only the
fixed point of the noncompact theory is available, and the spinons are deconfined. This result
confirms the analysis of Ref. \onlinecite{Hermele}, where it was argued that spinons deconfine
above a large but finite value of $N$. Thus, for $N>36$ the
theory is well controlled by an expansion
in $1/N$. For $N\in(20,36]$, on the other hand, it seems to correspond to an expansion
in the fugacity.
In order to investigate this claim, let us consider Eq. (\ref{f})
in the limit $l\to\infty$, in which case $f\to f_\pm=4\pi\varepsilon_\pm/N$, where
$\varepsilon_\pm=\lim_{l\to\infty}\varepsilon(ae^l)$. From Eqs. (\ref{f+-}) and (\ref{y+-}),
we obtain for $20<N\leq 36$
\begin{equation}
\varepsilon_\pm=\frac{N}{20}\left(1+\frac{1}{\pi}y^2_\pm\right),
\end{equation}
and
\begin{equation}
f_\pm=\frac{\pi}{5}\left(1+\frac{1}{\pi}y^2_\pm\right).
\end{equation}
The above equations clearly have the structure of an expansion in $y$. This
provides a further argument
for setting $\varepsilon=1$ in
Eqs. (\ref{betaf}) and (\ref{betay}). However,
it should be noted that within the present approximation 
$y^2_\pm$ is not small in the entire interval $20<N\leq 36$, and
perturbation theory may eventually breaks down
for some values of $N$.
Most critical is the situation at the fixed point
$(f_+,y_+)$, since $y_+$ gets small only for $N$ close to $36$. Note that also $f_+$ can be large
in this interval. Already at $N=20$ we have $f_+=\pi$, indicating that the
behavior near the fixed point $(f_+,y_+)$ should be considered with great care. The fixed point
$(f_-,y_-)$, on the other hand, has much better asymptotics. Indeed, $y_-$ is small for all values
of $N$ in the interval $20<N<36$ up to $N=32$, becoming larger than unity only above this value.

For all $20<N\leq 36$ the fixed points $(f_\pm,y_\pm)$ govern a confined
regime such that the interspinon potential at these fixed points reads
\begin{equation}
\label{V+-}
V_\pm(R)=\sigma_\pm R-\frac{f_\pm}{\pi^2 R}+V_L(R)+{\cal O}(1/R^2),
\end{equation}
where $\sigma_\pm$ is the string tension in the presence of
fermionic matter. In the present approximation, the string tension has precisely the
same form as in the $\overline{\rm QED}_{2,1}$,\cite{Polyakov-book} except that the bare parameters are
replaced by the renormalized ones. Thus, $\sigma=2e^2M/\pi^2=4e\sqrt{2z}/\pi$, or in
terms of dimensionless quantities, $\sigma/e_0^4=4\sqrt{2fy}/\pi$.
The term $V_L(R)$ is assumed to be given by a string
model for the electric flux tube due to L\"uscher,\cite{Luescher} which in
$d$ dimensions has the form
\begin{equation}
\label{VL}
V_L(R)=-\frac{(d-2)\pi}{24R}.
\end{equation}
Such a term  has to be included in order to account for the fluctuations of the electric flux tube
occuring even in absence of matter. The factor $d-2$ represents the number of transverse
modes of the string worldsheet. The coefficients of the $1/R$ terms in Eq. (\ref{V+-}) are universal and the string
tensions $\sigma_\pm$ reduce to the one obtained by Polyakov\cite{Polyakov} in the limit case where
no matter fields are present.

The schematic
flow diagram for $20<N<36$ is shown in Fig. 1.
In this case the system can be either in the confinement or
deconfinement phase, since the fixed points on each side of the dashed line in Fig.1 are
infrared stable. The fixed point $(f_-,y_-)$ is infrared stable only
along the dashed line in Fig. 1. Precisely
for this reason it plays an important role in our discussion. The point is that
$y_-$ vanishes for $N=20$, which is the value of $N$ leading to $\varepsilon_*=1$.
This result gives a string tension $\sigma_-$ that vanishes for all
$N\leq 20$, thus leading to a collapse of the fixed point $(f_-,y_-)$ into the fixed point
of the noncompact theory. Since the fixed point $(f_-,y_-)$ is unstable in the direction leading
to the fixed point $(f_+,y_+)$ and the one of the noncompact theory,
the latter will no longer be
stable and only the confinement phase governed by the infrared stable fixed point
$(f_+,y_+)$ will exist. Note that while the string tension $\sigma_-$ vanishes continously
as $N$ approaches $20$ from above, this is not the case as $N$ approaches 36 from below.
In fact, $\sigma_\pm/e_0^4=4\sqrt{2f_\pm y_\pm}/\pi$ has a finite value for $N=36$. Since
$\sigma$ vanishes for $N>36$, it follows that there is a universal jump in the string tension
for $N=36$.

It should be emphasized here that the physical interspinon potential corresponds
to the one associated with the infrared stable fixed point, i.e., $V_+(R)$. Since the fixed
point  $(f_-,y_-)$ is unstable, the dashed line represents a critical line separating
confined and deconfined regimes. This role of a separatrix played by the dashed line of
Fig. 1 is most clearly seen as $N$ cross the value $N=20$, in which case the line collapses
along the axes $f$ and $y$, with the fixed point  $(f_-,y_-)$ becoming identical to the
one of the noncompact theory.

The phase of the system when $20<N<36$ is determined by the size of
the Debye-H\"uckel parameter, $\kappa_D=n\lambda_D^3=\sqrt{2}e^2/(8\pi^2M)$,
where $n$ is the instanton density and
$\lambda_D$ the Debye length.
Since for $N>36$ only the fixed point at zero fugacity exists, we have
that $\sigma_+=0$, i.e., deconfinement occurs for $N>36$ and
the large-$N$ result is essentially correct.\cite{Hermele}
\begin{figure}
\includegraphics[width=7cm]{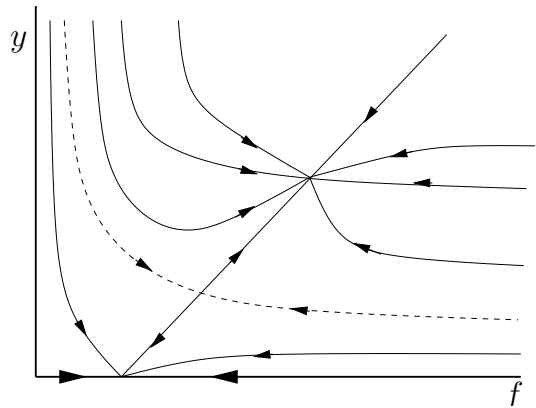}
\caption{Schematic flow diagram for $20<N<36$ featuring the fixed points at nonzero
fugacity given by Eqs. (\ref{f+-}) and (\ref{y+-}).}
\end{figure}

In Table I we summarize the phase structure of QSD.

\begin{table}
\caption{Summary of the phase structure of QSD. The third column
gives the universal coefficient of $1/R$.}
\begin{ruledtabular}
\begin{tabular}{ccc}
$N$ & Phase & $-1/R$ coefficient\\
\hline
$N\leq 20$ & confined & $\pi/24+f_+/\pi^2$ \\
$20<N\leq 36$ & confined/deconfined & like $N\leq 20$ or $N>36$ \\
$N>36$ & deconfined & $4/\pi N$
\end{tabular}
\end{ruledtabular}
\end{table}

\section{Discussion}

We have presented in this paper a
 description of
the quantum spinon dynamics
in a Heisenberg antiferromagnet 
 using a field theory
of $N$
identical replica of fermions
with a constraint.
Only the limit $N\rightarrow \infty$ can be treated exactly, the lower $N$ results
are obtained from different approximations.
Since $N=2$ corresponds to the
physically interesting case,
one may wonder
to which extent our results
are physically relevant. This
 question is
hard to answer. Originally, the study of spin liquids in 2+1 dimensions
was motivated by the properties of high-$T_c$ cuprate superconductors.\cite{Anderson}
It was hoped that doping a Mott insulator
provides a mechanism for
superconductivity in the cuprates. At zero doping the cuprates are antiferromagnetic
insulators, i.e., the global SU(2) spin symmetry
of the model is broken in the ground state. Spin
liquids, on the other hand, have no broken symmetries. Nevertheless, by doping a
spin liquid we expect to obtain a superconducting state.\cite{Wen-RMP,Senthil-Lee}
The idea
was that doping may frustrate the magnetic state of the system and produce a
spin liquid. With further doping the system would eventually become superconducting through
a kind of hole condensation. Hence doping plays a role similar to frustration
due to the triangular lattice in the organic compound $\kappa$-(ET)$_2$Cu$_2$(CN)$_3$,
where a spin liquid phase was recently reported.\cite{Kanoda}

Our analysis
 considers neither doping nor triangular lattices. Instead, the spin liquid emerges from
the large-$N$ limit, which provides the needed frustration to produce a spin
liquid. Our treatment at lower
values of $N$ was based
on an expansion in the instanton fugacity.
This treatment is essentially non-perturbative, and as such difficult to
control. Nevertheless the stability of the spin liquid for
large $N$ seems to be established beyond doubt.

Note, however, that the large-$N$
analysis of the lattice model is made in a particular representation of the
SU(N) group, by writing the spin operators in terms of fermions.\cite{Affleck,Arovas}
If the spin operators are written in terms of bosons --- which is an equally valid
description of the problem --- other representations are obtained. In such a case
the Berry phase plays an essential role in determining the phases of
the system.\cite{Read-Sachdev} There $N$ is kept fixed and large, while a
coupling constant $g$ related to the exchange constant is varied. 
In the lower phase, i.e.,
for $g<g_c$, where $g_c$ is the critical coupling,
the SU(N) symmetry is broken and we have a N\'eel-like state.
For $g>g_c$, on the other hand, the
paramagnetic phase {\it is not} a spin liquid. Instead, the so-called valence-bond
solid (VBS) state emerges.\cite{Read-Sachdev} For $N=2$ the VBS state can only
occur if frustrating interactions respecting the SU(2) symmetry are included in
the Hamiltonian. In such a scenario, where also an
effective compact U(1) gauge theory is featured, the spinons are confined in both
phases, but conjectured to be deconfined precisely at $g=g_c$. This is the so
called {\it deconfined quantum criticality scenario} and provides
a new paradigm for quantum phase transitions.\cite{Senthil-2004} One of the main
characteristics of the this new type of quantum criticality is a large value for
the $\eta$ exponent. In which systems this new paradigm actually holds is still under discussion.
There is a recent numerical evidence\cite{Sandvik} supporting the scenario proposed
in Ref. \onlinecite{Senthil-2004}. On the other hand, it was recently demonstrated using
large scale Monte Carlo simulations\cite{MC-easyplane} that quantum antiferromagnets
with easy-plane anisotropy do not exhibit deconfined quantum criticality,
contrary
 claims in Ref. \onlinecite{Senthil-2004}.

In QSD, our analysis certainly breaks down for $N=2$. The description in
terms of the generalized sine-Gordon theory (\ref{SG-new}) no
longer holds. The
reason for this is that the derivation of the Lagrangian  (\ref{SG-new}) assumes
that the fermions are massless,\cite{KNS,KNS1,Herbut1}
and it is known that
in noncompact QED$_{2,1}$ the fermion mass is dynamically generated through
chiral symmetry breaking (CSB) for small  $N$.\cite{Appelq-1,Appelq-2} While the precise
value of $N$ below which CSB occurs is still a matter of debate, it seems to be more or less
consensual that the chiral symmetry is broken for $N=2$.\cite{Appel} Massive fermions modify
the vacuum polarization in such a way that makes the derivation of an effective instanton Lagrangian
difficult.
Furthermore, independent of the CSB occuring in noncompact QED$_{2,1}$, confinement
is likely to
induce CSB in QSD already for $N$ below and near 20, in a regime where the effective Lagrangian
(\ref{SG-new}) can still be considered to be valid.
In any case, CSB corresponds to spin density waves and
this is precisely the state one would expect for undoped cuprates. The conclusion seems to be that no
spin liquid state is possible for $N=2$ in the undoped system, unless frustrating interactions are added.
In this case it may well be possible that a VBS emerges instead a spin liquid.
Moreover, the resulting
paramagnetic state may depend on the nature of the frustration. Here it is worth to mention that a 
local SU(2) gauge theory of (fermionic) spinons for 
a frustrated Heisenberg antiferromagnet\cite{Mudry-2d} exhibits a stable spin liquid phase.   

When doping is included the RG analysis becomes considerably
more difficult due to the coupling
of the gauge field with
a
non-relativistic complex scalar field.\cite{Wen-RMP} In this case there are additional topological
defects. These are
vortex excitations coupled to the instantons.
We are currently investigating this situation.\cite{NK-to-be-pub}

\appendix \section{Renormalization group equations for the
$d$-dimensional Coulomb gas}

In order to make the paper self-contained we  consider here the
general $d$-dimensional Coulomb gas, whose RG equations
were set up by Kosterlitz.\cite{Kosterlitz}
They were originally derived
using the  poor-man scaling approach.
Here we employ
the method due to Young,\cite{Young} which is
physically appealing, since it amounts to applying
a scale-dependent
Debye-H\"uckel argument which leads to the same results. Although
Young applied the method to derive the RG equations associated
to the Kosterlitz-Thouless (KT) phase transition, it
can easily be
generalized
to the $d$-dimensional case. We have done this
 previously \cite{KNS1}
to derive the RG equations for anomalous Coulomb gases in $d$-dimensions.
Here we concentrate on the ordinary $d$-dimensional Coulomb gas.

The bare Coulomb interaction between two opposite charges
of magnitude $\sqrt{K_0}$ is given by

\begin{equation}
\label{bareU}
U_0(r)=-4\pi^2K_0V(r),
\end{equation}
where

\begin{equation}
V(r)=\frac{a^{2-d}}{4\pi^{d/2}}\Gamma\left(\frac{d}{2}-1\right)
\left[\left(\frac{r}{a}\right)^{2-d}-1\right].
\end{equation}
In the above equation, $a$ is a short-distance cutoff, which
for $d=3$ will be set to $a=1/e_0^2$. From Eq. (\ref{bareU})
we obtain the bare electric field:

\begin{equation}
\label{bareE}
E_0(r)=-4\pi^2c(d)\frac{K_0}{r^{d-1}},
\end{equation}
where

\begin{equation}
c(d)=\frac{d-2}{4\pi^{d/2}}\Gamma\left(\frac{d}{2}-1\right).
\end{equation}

Next, in the spirit of the Debye-H\"uckel theory, we introduce
an effective medium via a scale-dependent dielectric constant
$\varepsilon(r)$. This gives the renormalized electric field
\begin{equation}
\label{renE}
E(r)=-4\pi^2c(d)\frac{K_0}{\varepsilon(r)r^{d-1}}.
\end{equation}
The dielectric constant
$\varepsilon(r) $ is expressed in terms of the
the susceptibility $\chi(r)$  as
\begin{equation}
\label{diel-const}
\varepsilon(r)=1+S_d\chi(r),
\end{equation}
where
\begin{equation}
\label{suscep}
\chi(r)=S_d\int_a^r ds s^{d-1}\alpha(s)n(s),
\end{equation}
with
 $S_d=2\pi^{d/2}/\Gamma(d/2)$ being
the surface of the
unit sphere in $d$ dimensions, and
 $\alpha(r)$ is the polarizability.
For
small separation of a dipole pair, it is given approximately by
\begin{equation}
\alpha(r)\approx\frac{4\pi^2K_0r^2}{d}.
\end{equation}
The average number of dipole pairs is
\begin{equation}
n(r)\approx z_0^2e^{-U(r)},
\end{equation}
where $z_0$ is the bare fugacity and $U(r)$
the renormalized potential obtained by integrating the renormalized
electric field (\ref{renE}),

\begin{equation}
\label{renU}
U(r)=U(a)+4\pi^2c(d)K_0\int_a^r\frac{ds}{s^{d-1}\varepsilon(s)}.
\end{equation}
The renormalized version  of $K_0$ is
\begin{equation}
\label{renK}
\frac{1}{K(l)}=\frac{\varepsilon(ae^l)}{K_0}e^{(d-2)l},
\end{equation}
where $l=\ln(r/a)$.
Differentiating Eq. (\ref{renU}) with respect to $l$,
and using
Eq. (\ref{renK}),
yields
\begin{equation}
\label{du}
\frac{dU}{dl}=\frac{4\pi^2c(d)}{a^{d-2}}K(l),
\end{equation}
Next we differentiate Eq. (\ref{renK}) with respect to $l$ to obtain

\begin{equation}
\label{prepRG-K}
\frac{dK^{-1}}{dl}=\frac{8\pi^2S_d^2z_0^2a^{d+2}}{d}e^{2dl-U(ae^l)}+(d-2)K^{-1}.
\end{equation}
Here
we define
\begin{equation}
\label{z2}
z^2(l)=\frac{8\pi^2S_d^2z_0^2}{d}e^{2dl-U(ae^l)},
\end{equation}
such that Eq. (\ref{prepRG-K}) becomes
\begin{equation}
\label{RG-K}
\frac{dK^{-1}}{dl}=a^{d+2}z^2+(d-2)K^{-1}.
\end{equation}
From Eq. (\ref{z2}) we derive the RG equation for the effective
fugacity:
\begin{equation}
\label{dz}
\frac{dz}{dl}=\left[d-\frac{2\pi^2c(d)K}{a^{d-2}}\right]z.
\end{equation}
It is convenient to introduce the dimensionless
quantities
$\kappa\equiv a^{2-d}K$ and $y\equiv a^d z$ to rewrite
Eqs. (\ref{RG-K}) and (\ref{dz}) as
\begin{eqnarray}
\frac{d\kappa^{-1}}{dl}&=&y^2+(d-2)\kappa^{-1},
\label{RG-K-new}
\\
\label{dz-new}
\frac{dy}{dl}&=&\left[d-2\pi^2c(d)\kappa\right]y.
\end{eqnarray}
~\\For $d=2$, the above RG equations govern the scaling behavior of the KT
transition, while for $d>2$ there is no fixed point, implying that
the $d$-dimensional Coulomb gas is always in the metallic phase.\cite{Kosterlitz}


\end{document}